\documentclass[11pt]{article}
\usepackage{amsmath}
\usepackage{graphicx}
\usepackage{enumerate}
\usepackage{natbib}
\usepackage{url} 
\usepackage{bm}
\usepackage{xcolor}
\usepackage{ltablex}
\usepackage{booktabs} 
\usepackage{longtable}
\usepackage{comment}
\usepackage[normalem]{ulem} 

\newcommand{\blind}{1}
\usepackage[bottom=1.2in,top=1in]{geometry}
\begin{document}

\def\spacingset#1{\renewcommand{\baselinestretch}%
{#1}\small\normalsize} \spacingset{1}


\if1\blind
{
  \title{\bf Bivariate Hierarchical Bayesian Model for Combining Summary Measures and their Uncertainties from Multiple Sources}
  \author{Yujing Yao$^{1}$, R.\ Todd Ogden$^{1}$, Chubing Zeng$^{2}$, and Qixuan Chen$^{1}$
    \hspace{.2cm}\\
    
    $^{1}$Department of Biostatistics, Mailman School of Public Health, \\
    Columbia University, New York, NY\\
    $^{2}$Division of Biostatistics, Department of Preventive Medicine, \\
    University of Southern California, Los Angeles, CA}
  \maketitle
} \fi

\if0\blind
{
  \bigskip
  \bigskip
  \bigskip
  \begin{center}
    {\LARGE\bf Bivariate Hierarchical Bayesian Model for Combining Summary Measures from Multiple Sources}
\end{center}
  \medskip
} \fi

\bigskip
\begin{abstract}
    It is often of interest to combine available estimates of a similar quantity from multiple data sources. When the corresponding variances of each estimate are also available, a model should take into account the \textit{uncertainty of the estimates} themselves as well as the \textit{uncertainty in the estimation of variances}. 
    In addition, if there exists a strong association between estimates and their variances, the correlation between these two quantities should also be considered. In this paper, we propose a bivariate hierarchical Bayesian model that jointly models the estimates and their estimated variances assuming a correlation between these two measures. 
    We conduct simulations to explore the performance of the proposed bivariate Bayesian model and compare it to other commonly used methods under different correlation scenarios. The proposed bivariate Bayesian model has a wide range of applications. We illustrate its application in three very different areas: PET brain imaging studies, meta-analysis, and small area estimation. 
\end{abstract}

\noindent%
\vfill

\spacingset{1} 

\section{Introduction}
\label{s:intro}

    In many areas of health and science research, it is common that multiple studies are conducted to address a similar question, or that similar measurements are available from data  collected in multiple geographic areas. Any estimate from a single study or geographic area may be affected by small sample size, missing data, or measurement error. In such a situation, inference can often be improved by combining estimates from the different studies or areas. To generalize treatment we use the term ``source'' to refer to each data origin from which an estimate is obtained. This general idea of combining estimates from multiple sources has wide applications. For example, meta-analysis involves combining the results of multiple independent studies \citep{glass1976primary, borenstein2011introduction}. Small area estimation in survey sampling involves the estimation of parameters in small areas by combining estimates of all areas to improve precision \citep{ghosh1994small,pfeffermann2002small,rao2015small}. 
    
    When data are comparable across sources, a natural approach to combine the estimates is simply to calculate their average. When data from the various sources have rather different characteristics, however, we can use instead a weighted average. The weights could be determined based on estimates of precision, sample sizes of sources, and other factors \citep{cochran1954combination, borenstein2011introduction}. However, if there is also uncertainty in the factors needed to compute the weights, it can lead to increased uncertainty of the resulting estimate. In such a case, weight trimming could be applied to reduce large weights to a maximum value, reducing variability but increasing bias \citep{potter1988survey, potter1990study}.
    
    Hierarchical models can be used as an alternative approach to combine estimates from multiple sources while accounting for heterogeneity between sources, e.g., variation in estimates \citep{browne2006comparison,goldstein2011multilevel}. In meta-analysis, a normal-normal hierarchical model is often used to combine estimates from individual studies to obtain a joint estimate, based on the assumption that the studies may be estimating distinct, but related effects \citep{dumouchel1994hierarchical,sutton2001bayesian,higgins2002quantifying,higgins2009re}. Similarly in small area estimation, the Fay-Herriot area-level model is a widely used hierarchical model for estimating parameters in small areas of a large survey by combining the unstable direct survey estimates of all small areas \citep{fay1979estimates,wang2003mean, rao2015small}. Both models consider the setting with a single outcome measure and assume the estimated variance of the summary measure to be a fixed quantity.
    
    More recently, hierarchical models are further developed in both meta-analysis and small area estimation for broader applications. In meta-analysis, \cite{reitsma2005bivariate} extended the normal-normal model to a bivariate hierarchical model exclusively for sensitivity and specificity in the area of diagnostic studies, assuming a bivariate normal distribution for logit sensitivity and logit specificity. Similar to the normal-normal hierarchical model, the variance estimates of sensitivity and specificity were assumed to be fixed values. To allow sparse data in the number of true positives, true negatives, false positives, or false negative in a study, \cite{chu2006bivariate} further developed a bivariate hierarchical model by assuming binomial distributions for the number of true positives and the number of true negatives and a hierarchical bivariate normal distribution for logit sensitivity and logit specificity \citep{paul2010bayesian, guo2017bayesian}. On the other hand, in small area estimation, the Fay-Herriot model was extended by taking into consideration the variability in the estimated sampling variance. Specifically,  \cite{you2006small} assumed the estimated variance has a chi-squared distribution.  \cite{maiti2014prediction} and \cite{sugasawa2017bayesian} modified the model by further assuming a underlying inverse-Gamma distribution of the true variance.

	Despite these new developments, the existing hierarchical models for summary measures still have some limitations. Both the univariate and bivariate hierarchical models in meta-analysis assume the observed variability of the summary measure to be fixed values \citep{reitsma2005bivariate, chu2006bivariate, fay1979estimates}. Although the uncertainty in the estimated variance is modeled in the modified Fay-Herriot models, independence was assumed between the estimated summary measure and its corresponding estimated variance \citep{wang2003mean,rao2015small,you2006small,maiti2014prediction,sugasawa2017bayesian}. However, variance of the summary measure in a study is usually unknown and estimated and thus the uncertainty in the variance estimate should also be modeled if it is not negligible. Further, there often exists a strong association between the observed summary measure and its corresponding variance estimate in a study, e.g. when the summary measure is a log odds ratio or a log rate. Neglecting the dependence between summary measures and their variances could result in a poor estimation of population and source-specific parameters. Motivated by this, we propose a bivariate hierarchical Bayesian model that jointly model a summary measure and its variance while allowing correlation between these two quantities both in their estimates and in the underlying parameters. This bivariate hierarchical model has wide applications in combining summary measures and their uncertainties from multiple sources, including but not limited to meta-analysis and small area estimation.

\section{Methods}
\label{s:model}

    We consider the situation in which we have estimates of similar measures from $n$ separate data sources. Let $y_{i}$ and $s_i$ denote the direct estimate of a summary measure and its uncertainty, respectively, from the $i^{th}$ source, and $\theta_i$ and $\sigma_i$ denote the corresponding source-specific true value of the parameters,  $i=1,\ldots, n$. Our initial goal is to obtain estimates of population averages denoted by $\mu$ across all $n$ sources and then also to obtain refined estimates of each $\theta_i$.  Building on this simplest case, we will then generalize the model to allow for the estimation of regression coefficients in a linear regression framework. 
    
    We first review the univariate hierarchical Bayesian model that has been widely used in meta-analysis and small area estimation in Section \ref{s:ubm}, and then present the new bivariate hierarchical Bayesian model in Section \ref{s:bbm}, in which we account for not only the uncertainty in estimating $s_i$ but also for the correlations between $y_i$ and $s_i$ and between $\theta_i$ and $\sigma_i$.

\subsection{A review of univariate hierarchical Bayesian model}\label{s:ubm}

    In a typical univariate hierarchical Bayesian model (UBM) for combining summary measures, $y_i$ is assumed to be normally distributed centered at the source-specific true value $\theta_i$ with variance $\sigma_i^2$. The source-specific values (the $\theta_i$'s) are then assumed to be normally distributed with a common mean $\mu$ and a common variance $\tau^2$. Specifically, 
    \begin{equation}\label{m:ubmnoCov}
        \begin{aligned}
            &f(y_i|\theta_i, \sigma_i) \sim N(\theta_i, \sigma_i^2) \\
            &f(\theta_i|\mu, \tau) \sim N(\mu, \tau^2),
        \,\, i=1,\ldots, n.
        \end{aligned}
    \end{equation}
    By assuming a uniform prior for $\mu$, when $\tau^2$ and $\sigma_i^2$'s are known, the posterior distribution of $\mu$ can be obtained as:
        \begin{equation}
        \label{postdist:mu}
            f(\mu|\bm{y}, \bm{\sigma}, \tau) \sim N \left(\frac {\sum_{i=1}^n \omega_i y_i}{\sum_{i=1}^n \omega_i}, \frac{1}{\sum_{i=1}^n \omega_i} \right), \text{~with~}\omega_i = \frac{1}{\sigma_i^2+\tau^2},
        \end{equation}
    where $\bm{y}$ is the $n\times 1$ vector $(y_1, \cdots, y_n)^T$ and $\bm{\sigma}$ is the $n\times 1$ vector $(\sigma_1, \dots, \sigma_n)^T$.
    That is, the common mean $\mu$ can be estimated using the posterior mean 
        \begin{equation}\label{e:ubmnoCov}
            \hat{\mu} = \frac {\sum_{i=1}^n \omega_i y_i}{\sum_{i=1}^n \omega_i}.
        \end{equation}
    In contrast to the typical weighted average estimator of $\mu$ that assigns weight $1/\sigma_i^2$ to the observation $y_i$, the inclusion of $\tau^2$ in the expression for $\omega_i$ reduces the chance of having extreme weights with very small $\sigma_i^2$, especially when $\tau^2$ is large relative to $\sigma_i^2$.

    When source-specific covariates $\bm{x_i}^T=(x_{i1}, \cdots, x_{ip})$ are available, Model (\ref{m:ubmnoCov}) can be extended to incorporate these into a regression model, with $\bm{\beta} = (\beta_1, \cdots, \beta_p)^T$ being a $p \times 1$ vector of coefficients:
        \begin{equation}\label{m:ubmwCov}
        \begin{aligned}
            &f(y_i|\theta_i, \sigma_i) \sim N(\theta_i, \sigma_i^2) \\
            &f(\theta_i|\bm{\beta}, \tau) \sim N(\bm{x_i}^T \bm{\beta}, \tau^2), \,\, i=1, \ldots, n.
        \end{aligned}
        \end{equation}
    If the $\sigma_i^2$'s and $\tau^2$ are known, with a uniform prior for ${\bm{\beta}}$, we can derive the posterior distribution of ${\bm{\beta}}$:
        \begin{equation}
            f(\bm{\beta}|\bm{y}, \bm{X}, \bm{\sigma}, \tau) \sim N \left(\left(\sum_{i=1}^n \omega_i \bm{x}_i \bm{x}_i^T\right)^{-1} \left(\sum_{i=1}^n \omega_i \bm{x}_i y_i\right), \left(\sum_{i=1}^n \omega_i \bm{x}_i \bm{x}_i^T\right)^{-1} \right)
        \end{equation}
    where $\bm{X}$ is the $n\times p$ matrix of $(\bm{x}_1, \cdots, \bm{x}_n)^T$. We can then estimate $\bm{\beta}$ using the posterior mean of ${\bm{\beta}}$:
        \begin{equation}\label{e:ubmwCov}
       \hat{\bm{\beta}} = \left(\sum_{i=1}^n \omega_i \bm{x}_i \bm{x}_i^T\right)^{-1} \left(\sum_{i=1}^n \omega_i \bm{x}_i y_i\right).
        \end{equation}
    The UBM is equivalent to the normal-normal hierarchical model in meta-analysis  \citep{dumouchel1994hierarchical} and the Fay-Herriot model in small area estimation \citep{fay1979estimates} (where interest also lies in estimating source-specific means $\theta_i, i=1, \ldots, n$).
    The posterior distribution for $\theta_i$ can be written as:
        \begin{equation}
        f(\theta_i |  y_i, \bm{x}_i, \bm{\beta}, \sigma_i, \tau) \sim N\left(\gamma_i y_i + (1-\gamma_i) \bm{x}_i^T \bm{\beta}, \,\,\, \sigma_i^2\gamma_i\right), \text{~where~}\gamma_i = \frac{\tau^2}{\tau^2 +\sigma_i^2 },
        \end{equation}
    and the estimate of $\theta_i$ can be obtained as the posterior mean with $\bm{\beta}$ substituted by its estimator in (\ref{e:ubmwCov}):
        \begin{equation}\label{e:ubmtheta}
       \hat{\theta}_i = \gamma_i y_i + (1-\gamma_i) \bm{x}_i^T \hat{\bm{\beta}}, \,\, i=1,\ldots, n.
        \end{equation}
    
\subsection{Bivariate hierarchical Bayesian model} \label{s:bbm}
    We consider a bivariate hierarchical Bayesian model (BBM)
    to take into account both the estimate and its variance, as well as the correlation between the two quantities.
    In the simplest scenario without source-specific covariates, the model is 
        \begin{equation}
        \label{m:bbmnoCov}
        \begin{aligned}
            &\left. \left( \begin{array}{c} y_i \\ \log s_i \\ \end{array} \right) \right\vert \theta_i, \sigma_i, \rho_1, \sigma_{s_i} \sim N_2 \left( \left( \begin{array}{c} \theta_i \\ \log \sigma_i \\ \end{array} \right),
        \left( \begin{array}{cc} \sigma_i^2 & \rho_1\sigma_i\sigma_{s_i} \\\rho_1\sigma_i\sigma_{s_i} & \sigma_{s_i}^2  \\ \end{array} \right) \right),
 \\
            &\left. \left( \begin{array}{c} \theta_i \\ \log \sigma_i \\ \end{array} \right) \right\vert \mu_{\theta}, \mu_{\sigma}, r_{\theta}, \rho_2, r_{\sigma} \sim N_2 
        \left( \left( \begin{array}{c} \mu_\theta \\\mu_\sigma \\ \end{array} \right),
        \left( \begin{array}{cc} r_\theta^2 & \rho_2 r_{\theta}r_{\sigma}  \\\rho_2 r_{\theta}r_{\sigma} & r_\sigma^2  \\ \end{array} \right) \right), \,\, i=1, \ldots, n.
        \end{aligned}
        \end{equation}
    This differs from existing methods in the following aspects:
    (1) We incorporate the uncertainties in estimating both $y_i$ and $\log s_i$ into the model using the variance $\sigma_i$ and $\sigma_{s_i}$.
    (2) We model $y_i$ and $\log s_i$ as bivariate normal random variables and model $\theta_i$ and $\log \sigma_i$ also as bivariate normal in the second level, while existing hierarchical Bayesian models in small area estimation specify a different underlying distribution for $s_i^2$, e.g., Chi-squared distribution, inverse Gamma distribution. 
    (3) Given the bivariate normal setting, we further introduce parameters $\rho_1$ and $\rho_2$ to allow correlations between $y_i$ and $\log s_i$ and between $\theta_i$ and $\log \sigma_i$, respectively.
    
    The conditional distribution for $y_i$ given $\log s_i$ and the conditional distribution of $\theta_i$ given $\log \sigma_i$ are
    \begin{equation}
    \label{m:bbmcond}
    \begin{aligned}
          & y_i \vert \log s_i, \theta_i, \sigma_i, \rho_1,  \sigma_{s_i} 
            \sim N \left( \theta_i + \rho_1 \frac{\sigma_i}{\sigma_{s_i}}(\log s_i  - \log \sigma_i) , \sigma_i^2 (1-\rho_1^2) \right), \\
         & \theta_i \vert \log \sigma_i, \mu_{\theta}, \mu_{\sigma}, r_{\theta}, \rho_2, r_{\sigma} \sim N \left( \mu_{\theta} + \rho_2 \frac{r_{\theta}}{r_{\sigma}}(\log \sigma_i  -  \mu_{\sigma}), r_{\theta}^2 (1-\rho_2^2) \right), \,\, i=1, \ldots, n.
    \end{aligned}
    \end{equation}
    With a flat prior on $\mu_{\theta}$, if all other parameters are known, the posterior distribution of $\mu_{\theta}$ can be obtained as:
    \begin{equation}
        \label{postdist:mu_theta}
            f(\mu_{\theta}|\bm{y}, \log \bm{s}, \bm{\sigma}, \rho_1, \bm{\sigma_s}, \mu_\sigma, r_\theta, \rho_2, r_\sigma) 
            \sim N \left(\frac {\sum_{i=1}^n \xi_i \Tilde{y}_i}{\sum_{i=1}^n \xi_i}, \frac{1}{\sum_{i=1}^n \xi_i} \right)
        \end{equation}
    where $\xi_i = \frac{1}{\sigma_i^2 (1-\rho_1^2) + r_{\theta}^2 (1-\rho_2^2)}$, $\Tilde{y}_i = \left (y_i - \rho_2 \frac{r_{\theta}}{r_{\sigma}} (\log \sigma_i  -  \mu_{\sigma})- \rho_1 \frac{\sigma_i}{\sigma_{s_i}} (\log s_i  - \log \sigma_i)  \right )$, $\log \bm{s}$ is the $n\times 1$ vector $(\log s_1, \cdots, \log s_n)^T$, and $\bm{\sigma_s}$ is the $n\times 1$ vector $(\sigma_{s_1}, \dots, \sigma_{s_n})^T$.
    The population mean can be estimated using the posterior mean 
     \begin{equation}\label{e:bbmnoCov}
            \Tilde{\mu}_\theta = \frac {\sum_{i=1}^n \xi_i \Tilde{y}_i}{\sum_{i=1}^n \xi_i}. 
    \end{equation}
    When $\rho_1 = \rho_2 = 0$, the posterior mean $\Tilde{\mu}_\theta$ in (\ref{e:bbmnoCov}) reduces to $\hat{\mu}$ in (\ref{e:ubmnoCov}) obtained from the UBM. If either of the correlations $\rho_1$ and $\rho_2$ is non-zero, $\hat{\mu}$ is biased. 
    
    When source-specific covariates $\bm{x}_i, i=1, \cdots, n$ are available, we replace the constant means $\mu_{\theta}$ and $\mu_{\sigma}$ in (\ref{m:bbmnoCov}) with regressions on the covariates:
        \begin{equation}\label{m:bbmwCov}
        \begin{aligned}
        &\left. \left( \begin{array}{c} y_i \\ \log s_i \\ \end{array} \right) \right\vert \theta_i, \sigma_i, \rho_1, \sigma_{s_i} \sim N_2 \left( \left( \begin{array}{c} \theta_i \\ \log \sigma_i \\ \end{array} \right),
        \left( \begin{array}{cc} \sigma_i^2 & \rho_1\sigma_i\sigma_{s_i} \\\rho_1\sigma_i\sigma_{s_i} & \sigma_{s_i}^2  \\ \end{array} \right) \right), \\
      &\left. \left( \begin{array}{c} \theta_i \\ \log \sigma_i \\ \end{array} \right) \right\vert \bm{\beta}_\theta, \bm{\beta}_\sigma, r_\theta, \rho_2, r_\sigma \sim N_2 
        \left( \left( \begin{array}{c} \bm{x}_i^T \bm{\beta}_\theta \\ \bm{x}_i^T \bm{\beta}_\sigma \\ \end{array} \right),
        \left( \begin{array}{cc} r_\theta^2 & \rho_2 r_{\theta}r_{\sigma}  \\\rho_2 r_{\theta}r_{\sigma} & r_\sigma^2  \\ \end{array} \right) \right), \,\, i=1,\ldots, n
        \end{aligned}
        \end{equation}
    where $\bm{\beta}_\theta$ and $\bm{\beta}_\sigma$ are regression coefficients associated with $\bm{x}_i$ in predicting $\theta_i$ and $\log \sigma_i$, respectively. The covariates can also be different in the models for $\theta_i$ and $\log \sigma_i$. 
    Similar to model (\ref{m:bbmnoCov}), with a uniform prior for $\bm{\beta}_\theta$, if all other parameters are known,  we can derive the posterior distribution of $\bm{\beta}_\theta$:
        \begin{equation}
        \begin{aligned}
         & f(\bm{\beta}_\theta|\bm{y}, \bm{X}, \log \bm{s}, \bm{\sigma}, \rho_1, \bm{\sigma_s}, \bm{\beta}_\sigma, r_\theta, \rho_2, r_\sigma) \\
         & \sim N \left(\left(\sum_{i=1}^n \xi_i \bm{x}_i \bm{x}_i^T\right)^{-1} \left(\sum_{i=1}^n \xi_i \bm{x}_i \breve{y}_i\right), \left(\sum_{i=1}^n \xi_i \bm{x}_i \bm{x}_i^T\right)^{-1} \right)
        \end{aligned}
        \end{equation}
    where $\breve{y}_i = \left (y_i - \rho_2 \frac{r_{\theta}}{r_{\sigma}} (\log \sigma_i  -  \bm{x}_i^T \bm{\beta}_\sigma )- \rho_1 \frac{\sigma_i}{\sigma_{s_i}} (\log s_i  - \log \sigma_i)  \right )$.
    We can then estimate the regression coefficient $\bm{\beta}_\theta$ using the posterior mean:
       \begin{equation}\label{e:bbmwCov}
       \Tilde{\bm{\beta}}_\theta = \left(\sum_{i=1}^n \xi_i \bm{x}_i \bm{x}_i^T\right)^{-1} \left(\sum_{i=1}^n \xi_i \bm{x}_i \breve{y}_i \right).
       \end{equation}
    Further, the posterior distribution of source-specific mean, $\theta_i$, can be obtained as:
        \begin{equation}
        \begin{aligned}
        & f(\theta_i |  y_i, \bm{x}_i, \log s_i, \bm{\beta}_\theta, \sigma_i, \rho_1, \sigma_{s_i}, \bm{\beta}_\sigma, r_\theta, \rho_2, r_\sigma) \\
        & \sim N\left(\zeta_i (y_i-\rho_1 \frac{\sigma_i}{\sigma_{s_i}} (\log s_i  - \log \sigma_i) ) + 
       (1-\zeta_i) ( \bm{x}_i^T {\bm{\beta}}_\theta + \rho_2 \frac{r_{\theta}}{r_{\sigma}} (\log \sigma_i  -  \bm{x}_i^T \bm{\beta}_\sigma ) ), \,\,\, \zeta_i \sigma_i^2(1-\rho_1^2) \right)
        \end{aligned}
        \end{equation}
    where $\zeta_i = \frac{r_\theta^2(1-\rho_2^2)}{r_\theta^2(1-\rho_2^2) +\sigma_i^2(1-\rho_1^2) }$. 
    We estimate $\theta_i$ using the posterior mean with $\bm{\beta}$ substituted by its estimator in (\ref{e:bbmwCov}):
        \begin{equation}\label{e:bbmtheta}
       \Tilde{\theta}_i = 
       \zeta_i \left (y_i-\rho_1 \frac{\sigma_i}{\sigma_{s_i}} (\log s_i  - \log \sigma_i)\right ) + 
       (1-\zeta_i) \left ( \bm{x}_i^T \Tilde{\bm{\beta}}_\theta + \rho_2 \frac{r_{\theta}}{r_{\sigma}} (\log \sigma_i  -  \bm{x}_i^T \bm{\beta}_\sigma ) \right ), \,\, i=1,\ldots, n.
        \end{equation}
    When $\rho_1= \rho_2 = 0$, the source-specific posterior mean $\Tilde{\theta}_i$ in (\ref{e:bbmtheta}) reduces to $\hat{\theta}_i$ in (\ref{e:ubmtheta}) obtained from the UBM. If $\rho_1$ and $\rho_2$ are ignored when either of them is not equal to 0, the estimator $\hat{\theta}_i$ in (\ref{e:ubmtheta}) from the UBM can be biased even if we have good estimates for $\tau^2$ and $\sigma_i^2, i=1, \cdots, n$. 
    
    In some applications, the normality assumption may not be appropriate, e.g., when source-level estimates are proportions \citep{fabrizi2016hierarchical,sugasawa2018hierarchical}. In such a case, we consider applying some transformation $g(\cdot)$ to the outcome, and the variance of the transformed outcome can be approximated by $g^\prime(y_i)^2\sigma_i^{2}$.
    
    \subsection{Bayesian computation}
    To get full inference of the above models, Bayesian statistics can be used by specifying independent prior distributions for all non-intermediate parameters.
    To ease the computation, we apply Cholesky parameterization for the covariance matrix in the second level of the multivariate normal distribution. Specifically, the covariance matrix is decomposed as
    \begin{align*}
        \left( \begin{array}{cc} r_\theta^2 & \rho_2 r_{\theta}r_{\sigma}  \\\rho_2 r_{\theta}r_{\sigma} & r_\sigma^2  \\ \end{array} \right) = 
        \left( \begin{array}{cc} r_\theta & 0  \\0 & r_\sigma  \\ \end{array} \right) LL^T \left( \begin{array}{cc} r_\theta & 0  \\0 & r_\sigma  \\ \end{array} \right),
    \end{align*}
    with the Cholesky factor of the correlation matrix $L = \left( \begin{array}{cc} 1 & 0  \\ \rho_2 & \sqrt{1-\rho_2^2}  \\ \end{array} \right)$. Then we place an LKJ prior distribution on the Cholesky factor rather than placing a non-informative prior on $\rho_2$ \citep{gelman2006data}.
    
    Not all these parameters have a closed-form expression for their posterior distributions, and thus Gibbs sampling can be challenging. We use Stan's NUTS-Hamiltonian Monte Carlo (HMC) sampler, via RStan \citep{carpenter2017stan}, for the Bayesian computation of BBM. HMC is a Markov chain Monte Carlo (MCMC) method that uses the derivatives of the density function being sampled to generate efficient transitions spanning the posterior. It can properly explore high-dimensional target distributions, and is faster and more scalable \citep{neal2011mcmc,hoffman2014no, betancourt2015hamiltonian}.
    To check for sampling behavior and model convergence, we consider trace plots, the effective sample size, and the Gelman-Rubin diagnostic statistic $\hat{R}$ \citep{gelman1992inference}. 
    An R-package \texttt{bmsSum} for \textbf{B}ayesian \textbf{M}odel with \textbf{S}tan for combining \textbf{Sum}mary measures using UBM and BBM with or without covariates is available on GitHub.
    
\section{Simulation}\label{s:simu}

\subsection{Simulation Design} \label{s:simuset}
    We used the synthetic data and compared BBM with UBM and other existing methods to access how well they each estimate 
    \begin{enumerate}
        \item the population mean, i.e., $\mu$ in model (\ref{m:ubmnoCov}) and $\mu_{\theta}$ in model (\ref{m:bbmnoCov}) (section \ref{s:mu}),
        \item the regression coefficient, i.e., $\bm{\beta}$ in model (\ref{m:ubmwCov}) and $\bm{\beta}_{\theta}$ in model (\ref{m:bbmwCov}) (section \ref{s:beta}),
        \item the source-specific means, i.e., $\theta_i, i=1, \cdots, n$ in model (\ref{m:ubmnoCov}) and (\ref{m:bbmnoCov}) (section \ref{s:theta}).
    \end{enumerate}
    We considered four scenarios with different combinations of $\rho_1$ and $\rho_2$, where $\rho_1$ is the correlation between $y_i$ and $\log s_i$ and $\rho_2$ is the correlation between $\theta_i$ and $\log \sigma_i$ (see models in Section \ref{s:bbm}). Those scenarios are: (1) $\rho_1=\rho_2=0$; (2) $\rho_1 \neq 0, \rho_2=0$; (3) $\rho_1=0, \rho_2 \neq 0$; (4) $\rho_1=\rho_2 \neq 0$.
    
    In each simulation, we generated a random sample of $n=50$ observations. We used Rstan to fit both UBM and BBM with independent improper uniform priors for $\mu$, $\mu_{\theta}$ and $\mu_{\sigma}$; $\texttt{Normal}(0, 10^6)$ prior for $\beta_{\theta,j}$, and $\beta_{\sigma, j}$, $j=1, \ldots, p$; $\texttt{Half-Cauchy}(2.5)$ prior for $\tau$, $r_\theta$, $r_\sigma$, and $\sigma_{s_i}$; $\texttt{LKJCorr}(4)$ prior for the Cholesky factor $L$; and $\texttt{Unif}(-1,1)$ prior for $\rho_1$. To obtain the posterior distributions of parameters of interest,
    we ran three chains with 5000 iterations, 2000 warm-up and a lag of 10 in each chain, which generated 900 draws for each model parameter. 
    Point estimates are the means of the posterior distributions, and the 95\% credible interval (CI) was constructed by equally splitting the tail areas of the posterior distributions. 
    We generated $500$ datasets for each scenario. For each simulation setting and for each estimator, we calculated empirical bias, mean squared error (MSE), and coverage rate of the corresponding intervals.
    
\subsection{Estimation of population mean} \label{s:mu}
    To address the first question, we generated 
    $(\theta_i$, $\log \sigma_i)^T$, $i =1, \cdots, n$ from a bivariate normal distribution with mean vector $(\mu_\theta = 10, \mu_\sigma =2)^T$ and variance components $r_\theta =3, r_\sigma=1$, and $\sigma_{s_i} = 1$. The correlations, $\rho_1$ and $\rho_2$, can take values $0, 0.3, 0.5$ and $0.7$. To account for the situation when data sources are more homogeneous, we also generated data with smaller $\text{log}\sigma_i$ by using $\mu_{\sigma}=0.2$ and $r_{\sigma}=0.1$. 
    We compared the BBM estimator to the UBM estimator as well as the following three estimators from commonly used (non-Bayesian) methods, including 
        \begin{enumerate}
            \item (``raw''): $\frac{1}{n}\sum_{i=1}^n y_i$;
            \item (``weighted'') estimation with weights $1/s_i^2$: $\sum_{i=1}^n \frac{1}{s_i^2} y_i/\sum_{i=1}^n \frac{1}{s_i^2}$;
            \item (``trimmed'') weighted estimation with weights $\omega_i$ trimmed to 3 times the mean $1/s_i^2$: \\
            $\sum_{i=1}^n \omega_i y_i/\sum_{i=1}^n \omega_i$, where $\omega_i = 1/s_i^2$ if $1/s_i^2 \le \frac{3}{n}\sum_{i=1}^n 1/s_i^2$ and $\omega_i = \frac{3}{n}\sum_{i=1}^n 1/s_i^2$ if $1/s_i^2 > \frac{3}{n}\sum_{i=1}^n 1/s_i^2$ \citep{chen2017approaches}.
        \end{enumerate}
    For ``raw'' and ``weighted'' estimators, 95\% confidence intervals (CIs) were based on the normality assumption. For the ``trimmed'' estimator, 95\% CIs were based on bootstrap samples.

\begin{figure}
\begin{center}
\includegraphics[width=5.5in]{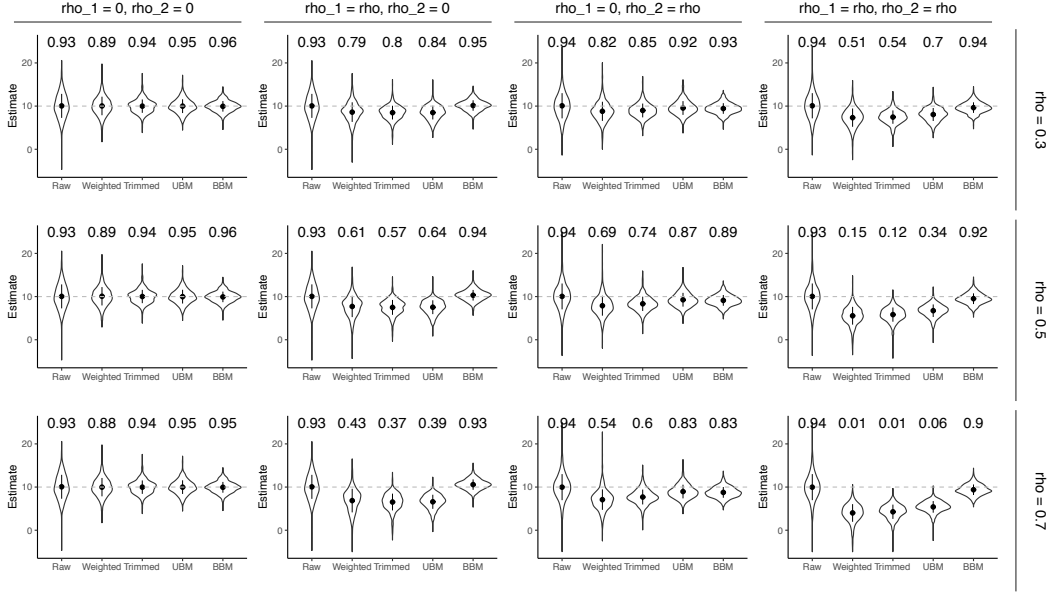}
\end{center}
\caption{Comparison of BBM and UBM to the three non-Bayesian methods for estimating the population overall mean with $\rho_1$ and $\rho_2$ taking different values. The violin plot presents the distribution of estimates and the number above the violin plot shows the 95\% CI coverage rate based on $500$ simulations.}
\label{f-1}
\end{figure}

        
    Results for the setting with more heterogeneity in the summary measures, $y_i$, across data sources (i.e., $\mu_{\sigma}=2$ and $r_{\sigma}=1$) are shown in Figure \ref{f-1} and Supplementary Table S1. When $\rho_1=\rho_2=0$, all methods provide unbiased estimates of the overall mean. However, the raw estimator and weighted estimator yield larger variation than the other three estimators;  the smallest variation is observed for the BBM estimator. All methods yield a coverage rate close to 95\%, except for the weighted estimator. The UBM works well in this scenario but not as well as the BBM, since the UBM uses $s_i$ in place of the true $\sigma_i$.  The BBM works well without over-fitting the data even though there are no correlations between measures and uncertainties of the measures in this scenario. 

    When at least one of $\rho_1$ or $\rho_2$ is nonzero, the BBM method outperforms the other estimators since it takes both types of correlations into consideration. Similar to the $\rho_1=\rho_2=0$ scenario, the raw estimator provides unbiased estimate with close to the nominal level coverage rate but displays rather large variation. The two weighted estimators and the UBM do not perform well in general. When $\rho_1 \neq 0$ and $\rho_2 = 0$, the two weighted methods, and UBM yield biased estimation with CIs below the nominal level coverage rate, and the bias and under-coverage becomes more severe for larger values of $\rho_1$.  When $\rho_1=0$ and $\rho_2 \neq 0$, the two weighted estimators still perform poorly but the UBM performs reasonably well. Finally, when both $\rho_1$ and $\rho_2$ are nonzero, the two weighted estimators and the UBM perform even worse with very large bias and very poor coverage. It's interesting to note that when $\rho_2 \neq 0$ as compared to the situation in which $\rho_1 \neq 0$, the UBM estimator yields less bias. The weighted estimator with trimmed weights can reduce variation in the estimate compared to the weighted estimator without weight trimming, but this step can also introduce bias and may lead to worse interval coverage.
    
    When data sources are less heterogeneous  with small variations between $y_i$ ($\mu_{\sigma}=0.2$ and $r_{\sigma}=0.1$), all methods perform better than the setting with more variations between $y_i$. BBM still performs better than UBM and the weighted estimators with or without trimming, especially when $\rho_1 \neq$ 0. The raw estimator performs similarly to BBM now with estimates centered at the true population mean, small variations in the estimates, and coverage rate close to 0.95 (see Supplementary Figure S1). 
        
\subsection{Estimation of regression coefficient} \label{s:beta}
    To address the second question, we generated the design matrix $\bm{X}$ of the regression model (\ref{m:bbmwCov}) with 3 columns, including a vector of 1 for intercept,  $\bm{x}_1 \sim \texttt{Normal}(0,1)$, and $\bm{x}_2 \sim \texttt{Bernoulli}(0.2)$. We set the regression coefficients $\beta_{\theta} = (5,3,1)^T$ and $\beta_{\sigma} = (1,1,0)^T$, with the same variance-covariance matrix and correlation setting in section (\ref{s:mu}). We compared the BBM estimator to the UBM estimator as well as to estimators obtained from three non-Bayesian approaches, including (unweighted) linear regression (LR), weighted linear regression with weights $1/s_i^2$ (WLR), and weighted linear regression with trimmed weights (TWLR) as defined above.

\begin{table}[!ht]
\caption{Comparison of bias, MSE, and coverage rate of 95\% CIs of the two Bayesian model-based estimators and the three non-Bayesian estimators in estimating the slope associated with the continuous predictor $\bm{x}_1$ under different combinations of $\rho_1$ and $\rho_2$.}
\label{t-2}
\begin{center}
\scalebox{0.8}{
\begin{tabular}{c|c|ccc|ccc|ccc}
\hline
\textbf{Correlation} & &
\multicolumn{3}{c}{\textbf{$\rho=0.3$}} & \multicolumn{3}{c}{\textbf{$\rho=0.5$}} &  \multicolumn{3}{c}{\textbf{$\rho=0.7$}} \\\hline
Scenario    & $\beta_{\theta,1}=3$      & Bias & MSE   & Coverage & Bias  & MSE   & Coverage & Bias  & MSE   & Coverage \\\hline
$\rho_1=0, \rho_2 = 0$  & LR   & 0.02  & 13.83 & 0.89 & 0.02  & 13.83 & 0.89 & 0.02  & 13.83 & 0.89 \\
         & WLR         & 0.02  & 5.14  & 0.41 & 0.02  & 5.14  & 0.41 & 0.02  & 5.14  & 0.41 \\
         & TWLR        & 0.03  & 2.85  & 0.54 & 0.03  & 2.85  & 0.54 & 0.03  & 2.85  & 0.54 \\
         & UBM         & 0.01  & 1.51  & 0.91 & 0.01  & 1.51  & 0.91 & 0.01  & 1.51  & 0.91 \\
         & BBM         & -0.02 & 0.74  & 0.94 & -0.02 & 0.74  & 0.94 & -0.02 & 0.74  & 0.94 \\\hline
$\rho_1 = \rho , \rho_2 = 0$  & LR  & 0.02  & 13.83 & 0.89 & 0.01  & 12.57 & 0.88 & 0.11  & 11.31 & 0.88  \\
         & WLR         & -0.29 & 5.06  & 0.42 & -0.52 & 5.12  & 0.42 & -0.67 & 5.46  & 0.4  \\
         & TWLR        & -0.3  & 2.9   & 0.53 & -0.57 & 3.17  & 0.52 & -0.77 & 3.5   & 0.51     \\
         & UBM         & -0.69 & 1.96  & 0.85 & -1.19 & 3.31  & 0.76 & -1.64 & 4.83  & 0.65 \\
         & BBM         & 0.03  & 0.72  & 0.95 & 0.06  & 0.71  & 0.95 & 0.06  & 0.68  & 0.95 \\\hline
$\rho_1 =  0, \rho_2 = \rho $   & LR   & 0.18  & 13.13 & 0.88 & 0.06  & 13.5  & 0.86 & 0.2   & 12.34 & 0.87 \\
         & WLR         & -0.1  & 5.21  & 0.45 & -0.2  & 5.22  & 0.39 & -0.41 & 5.03  & 0.38 \\
         & TWLR        & -0.14 & 2.98  & 0.52 & -0.28 & 2.98  & 0.53 & -0.49 & 3.06  & 0.48 \\
         & UBM         & -0.17 & 1.59  & 0.9  & -0.24 & 1.67  & 0.89 & -0.32 & 1.91  & 0.86 \\
         & BBM         & -0.11 & 0.67  & 0.94 & -0.16 & 0.68  & 0.95 & -0.24 & 0.67  & 0.96 \\\hline
$\rho_1 = \rho_2 = \rho $   & LR  & -0.01 & 12.37 & 0.89 & 0.15  & 13.14 & 0.88 & 0.21  & 12.95 & 0.87 \\
         & WLR         & -0.34 & 5.16  & 0.44 & -0.75 & 5.43  & 0.36 & -1.15 & 5.28  & 0.32 \\
         & TWLR        & -0.46 & 2.87  & 0.54 & -0.89 & 3.4   & 0.43 & -1.33 & 4.23  & 0.32 \\
         & UBM         & -0.88 & 2.2   & 0.82 & -1.47 & 4.15  & 0.64 & -2.03 & 6.47  & 0.39 \\
         & BBM         & -0.06 & 0.67  & 0.96 & -0.09 & 0.62  & 0.96 & -0.13 & 0.56  & 0.95 \\\hline
\end{tabular}}
\end{center}
\end{table}
    
    Simulation results are provided in Table \ref{t-2} for the slope of $X_1$ and in Supplementary Table S2 for the intercept and slope of $X_2$. The findings are similar to the estimation of population mean. The LR estimates are  slightly less biased but have much larger MSE and  nominal level coverage rates that are below the nominal rate for all scenarios. The BBM performs best, yielding the smallest bias and MSE with the coverage rate close to the nominal level. The UBM performs poorly with large bias,  large MSE, and below nominal level coverage rate when $\rho_1 > 0$, and its performance deteriorates for larger values of $\rho_1$. In contrast, the UBM performs reasonably well when $\rho_1 = 0$ even if $\rho_2 > 0$. The weighted estimator with trimmed weights is more efficient than the weighted estimator without weight trimming, and both are also more efficient than the LR estimator but less efficient than the two Bayesian model-based estimators. The 95\% CIs of the two weighted estimators yield very poor coverage rates.

\subsection{Estimation of source-specific means} \label{s:theta}
    In some applications like small area estimation, interest lies in estimating $\theta_i$, the mean of $Y$ from the $i^{th}$ data source. In such a situation we are interested in how the estimator of UBM and BBM improves over the ``raw'' estimate $y_i$. We considered the scenario in which there is correlation between a measurement and its uncertainty. As was done previously, we generated $(\theta_i, \log \sigma_i)^T, i =1, \cdots, n$ from bivariate normal distribution with mean vector $(\mu_\theta = 10, \mu_\sigma =2)^T$,  variance components $r_\theta =3, r_\sigma=1, \sigma_{s_i} = 1$, and set $\rho_1 = \rho_2 = 0.7$. One simulation was performed with a random sample of $n=20$ observations, and we compared the point estimates of $\theta_i$ and 95\%  CIs using the posterior distributions of parameters in BBM and UBM to the observed data $y_i$ and the true $\theta_i$.

\begin{figure}
\begin{center}
\includegraphics[width=4.5in]{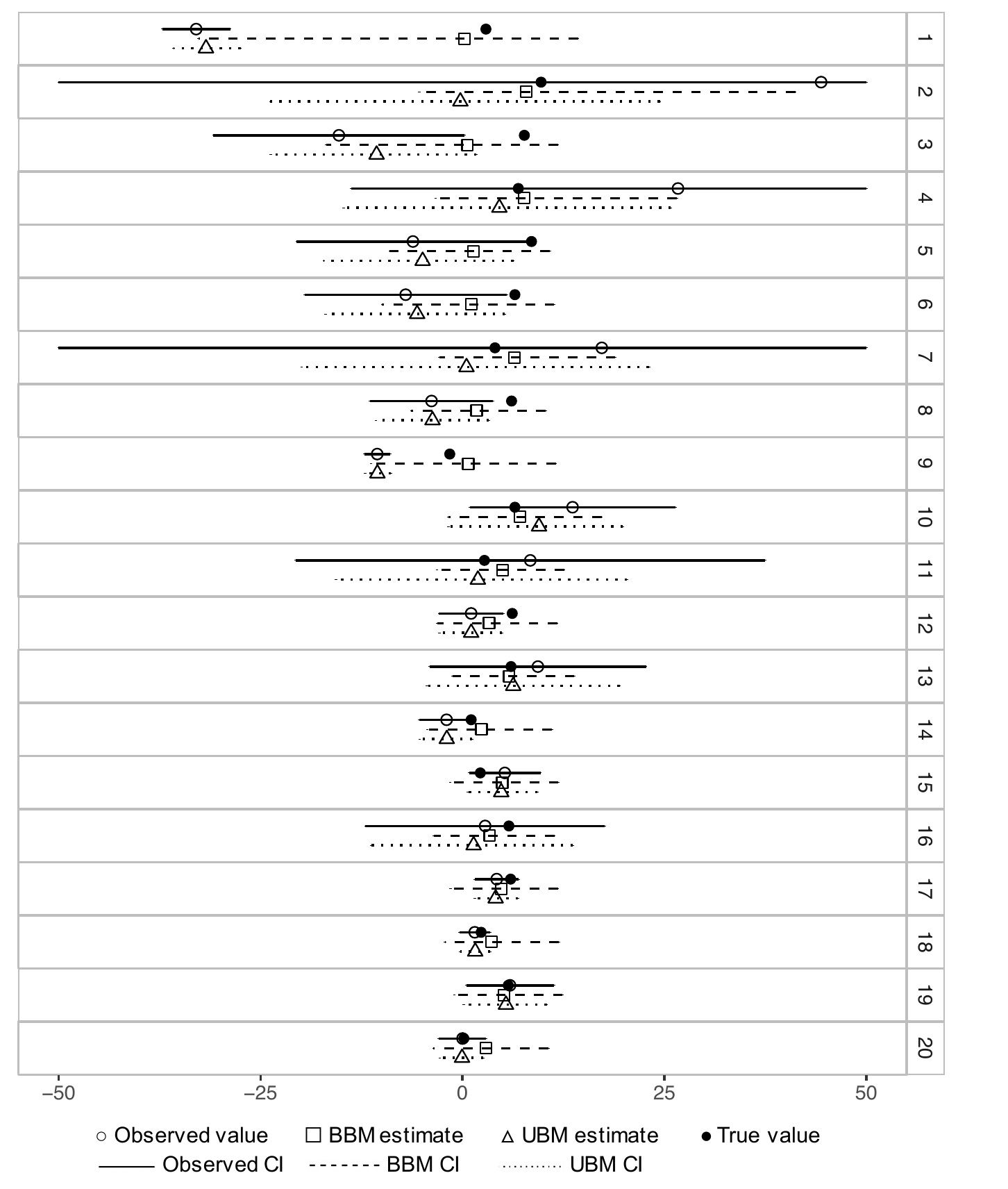}
\end{center}
\caption{The plot of $\theta_i$ (true value) , $y_i$ (observed value) with 95\% CI (top, solid line, some truncated at (-50, 50)), and comparison to the point estimates and 95\% CIs of BBM (middle, dashed line) and UBM (bottom, dotted line) for a simulated data with 
$n=20$. Results are sorted by the descending absolute distance of $\theta_i$ and $y_i$ . Numbers on the right column indicate the sorting order $i=1, \cdots 20$.}
\label{f-2-2}
\end{figure}

    Figure \ref{f-2-2} shows the estimates and 95\% CIs for estimating the source-specific mean $\theta_i$ using UBM, BBM and the raw estimate $y_i$, sorted by the absolute distance between $y_i$ and $\theta_i$. When $y_i$ is close to $\theta_i$ and the CI for $y_i$ is narrow (e.g., in the case $i = 17$), the estimates of UBM and BBM are similar but UBM yields a shorter 95\% CI. This indicates that when the direct estimate $y_i$ is already a good estimate of $\theta_i$, the simpler UBM tends to have better estimation performance than BBM. When $y_i$ is close to $\theta_i$ but $y_i$ has a wide 95\% CI (e.g., in the case $i = 11$), estimates of UBM and BBM are still close but BBM improves efficiency and thus yields a shorter CI. When $y_i$ is farther away from $\theta_i$, BBM performs much better than UBM. Specifically, the BBM estimate tends to be closer to $\theta_i$ than that of UBM; when the CI of $y_i$ is narrow and does not cover $\theta_i$ (e.g., in the case $i = 1$), BBM yields a wider CI that contains $\theta_i$ more frequently than UBM; when the CI of $y_i$ is wide and covers $\theta_i$ (e.g., in the case $i = 4$), BBM improves efficiency and yields a shorter CI while still containing $\theta_i$.

\section{Real data studies}\label{s:appl}

    We illustrated the application of BBM using three very different data examples, including  PET brain imaging, meta-analysis, and small area estimation.  In the applications, we assessed model fitting of BBM using the Bayesian posterior predictive $p$-value \citep{rubin1984bayesianly, gelman1996posterior,gelman2013bayesian}:
    \begin{align*}
    p = Pr \left (T(\bm{Z}_{rep},\bm{\Psi}) \geq T(\bm{Z}_{obs},\bm{\Psi} ) | \bm{Z}_{obs} \right ),
    \end{align*}
    where $\bm{Z} = (\bm{y}, \log \bm{s})$. Note that $T(\cdot)$ is a test statistic that depends on data $\bm{Z}$ and  parameters denoted using $\bm{\Psi}$. $\bm{Z}_{obs}$ denotes observed data and $\bm{Z}_{rep}$ denotes replicated data drawn from the posterior predictive distributions. If a model fits the data well, $T(\bm{Z}_{obs},\bm{\Psi})$ will be close to the center of the density plot of $T(\bm{Z}_{rep},\bm{\Psi})$. In other words, the posterior predictive $p$-value will be close to 0.5. Extreme $p$-values (near 0 or 1) suggest poor fit. Naturally, the choice of test statistic $T(\cdot)$ varies according to the application at hand \citep{crespi2009bayesian}. We used a test statistic that measures the discrepancy between the observed data $y_{obs,i}$ and the fitted distribution of $y_i$ given $\log s_i$ across all data sources $i=1, \cdots, n$:
        $$
            T(\bm{Z}_{obs},\bm{\Psi}) = \sum_{i=1}^n \frac{(y_{obs,i}-E(y_i|\log s_{obs,i}, \bm{\Psi}))^2}{Var (y_i|\log s_{obs,i}, \bm{\Psi})},
        $$
    where $E(y_i|\log s_{obs,i}, \bm{\Psi}) = \theta_i + \rho_1 \frac{\sigma_i}{\sigma_{s_i}}(\log s_i-\log\sigma_i)$ and $Var (y_i|\log s_{obs,i}, \bm{\Psi}) = \sigma_i^2(1-\rho_1^2)$ according to Model (\ref{m:bbmcond}). 

\subsection{Application to PET brain imaging data}\label{s:pet}

    In the study of the human brain,  positron emission tomography (PET) allows \emph{in vivo} measurement of the density of a protein of interest through modeling the kinetics of the concentration of a radioactive ligand over time \citep{morris2004kinetic,carson2005tracer}.  This is typically done separately for each subject, but interest generally lies in the population average. To estimate this average, subject-level estimates may be weighted according to estimates of precision which is also calculated at the subject level.  These can depend on both biological factors as well as the amount of injected dose, the presence of imaging artifacts, the noise level in measurements of blood samples necessary for quantification, etc.
    However, the resulting weighted estimate of a population mean can be unstable since the estimates of precision themselves are typically not very precise, and any underestimate of variance (arising purely due to chance) can result in extremely large weights. Since PET imaging is invasive, expensive, and labor intensive, sample sizes in PET studies are typically small, and so it is vitally important to combine all measures into a population estimate as efficiently as possible. 
    
    In our data set, 82 subjects, including 51 depressed subjects and 31 normal controls, were each scanned as part of a study examining the density of the serotonin transporter throughout the brain.  
    The subject-level estimate of the binding potential (a measure of the density of the transporters) is calculated based on the acquired sequence of PET images, along with measurements taken on blood samples drawn during the scan. Using a bootstrap algorithm \citep{ogden2005estimation}, it is possible to obtain an estimate of the variance of the estimated binding potential for each subject. Primary interest lies in investigating the population average of the binding potential and whether it differs on average between patients and control subjects.

\begin{figure}
\begin{center}
\includegraphics[width=4.5in]{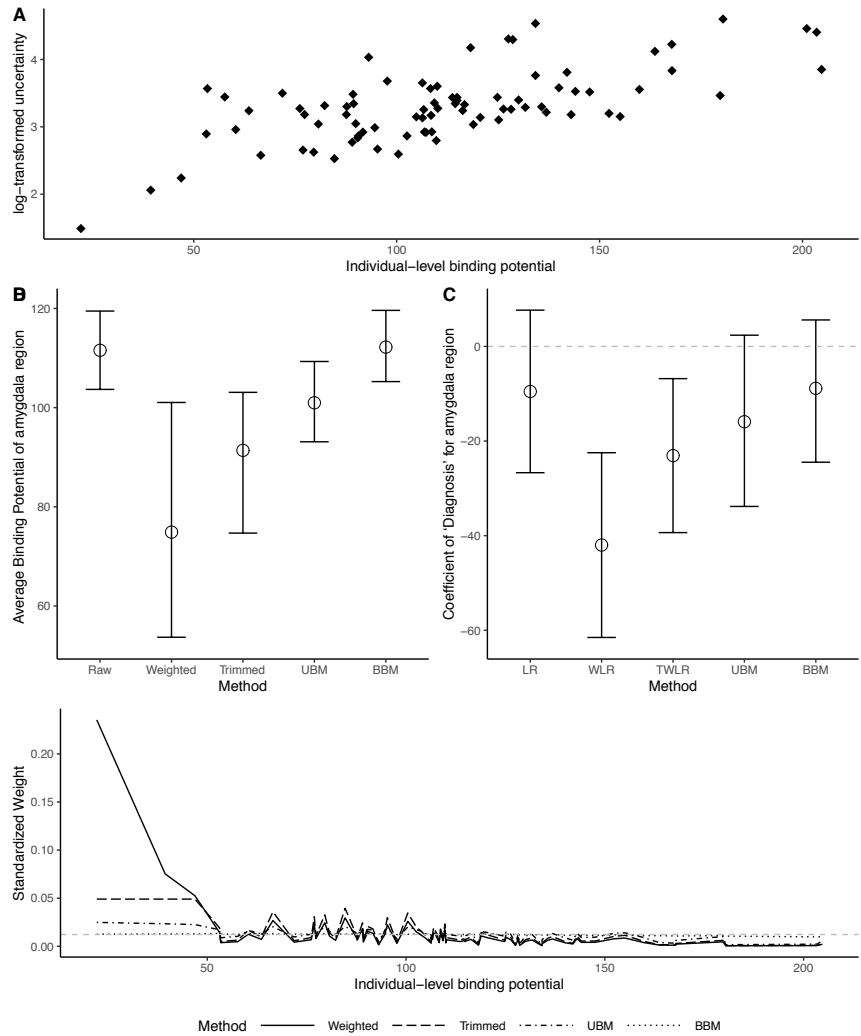}
\end{center}
\caption{Application to PET brain imaging data: (A) Scatter plot of log-transformed uncertainty (y-axis) versus  individual-level binding potential (x-axis) ; (B) Comparison of impact of each data point (measured using standardized weights) to the estimation of the overall binding potential mean in \texttt{amygdala} region using the weighted, trimmed, UBM and BBM methods. The grey dashed line at $y=1/82$ represents the setting of equal contribution from all data points; (C) Comparison of estimates and 95\% CIs for binding potential of \texttt{amygdala} region population average $\mu$ using different methods; (D) estimation of regression coefficient associated with diagnosis group.}
\label{f-3}
\end{figure}
    
    We illustrated the methods described in section \ref{s:simuset} by applying them to the PET imaging data, focusing on the amygdala region.
    The data suggest a positive correlation between the individual binding potential estimates and the corresponding log-transformed variance estimates  (Figure~\ref{f-3}(A)).  The estimated population average of the binding potential based on BBM is 112.2 (95\% CI: 105.3, 119.6), which is close to the result of unweighted method (Figure~\ref{f-3}(B)). The two weighted estimators lead to smaller estimates of the population average with wider 95\% CIs. By trimming extreme weights, the trimmed method provides a narrower CI. The UBM falls between the estimates of the weighted methods and the BBM with shorter CI than the weighted estimates. Figure~\ref{f-3}(C) shows the estimate of the regression coefficient associated with diagnosis group after adjusting for age and gender using corresponding regression models. The patterns are similar to the results for the population average estimation without any covariates, and BBM still provides the shortest 95\% credible interval among all the methods. The BBM shows an estimate of -8.9 (95\% CI: -24.5,5.6) for the coefficient associated with diagnosis group and suggests that the binding potential of amygdala region was not different between the two diagnosis groups. The conclusion is consistent with LR, UBM, but different from WLR and TWLR. The corresponding numerical results are in Supplementary Table S3. The Bayesian posterior predictive $p$-value is 0.47 and 0.49 in the BBM model without and with covariates, respectively, suggesting proper fit of the BBM models to the data (Supplementary Figure S2).
    
    To further investigate the differences of UBM, BBM compared to existing weighted methods in the analysis for the PET brain imaging data, we visualize the contribution of each data point in estimating the overall mean.  Figure~\ref{f-3}(D) shows the standardized weights (denoted by $\lambda_i$, such that $\sum_{i=1}^n \lambda_i =1$) of each data point using the different methods. Specifically, $\lambda_i=w_i/\sum_{i=1}^n w_i, i=1, \cdots, n$, with $w_i$ being $\frac{1}{s_i^2}$ for the weighted estimator, the trimmed weights for the trimmed method, $\frac{1}{\sigma_i^2+\tau^2}$ in Formula (2) for UBM, and $ \frac{1}{\sigma_i^2(1-\rho_1^2)+r_\theta^2(1-\rho_2^2)}$ in Formula (11) for BBM. 
    The parameters $\sigma_i^2$, $\tau^2$, $r_\theta^2$, $\rho_1$ and $\rho_2$ in weights of UBM or BBM are estimated by their posterior means.  
    The BBM in Model (\ref{m:bbmnoCov}) accounts for the correlation between $y_i$ and $s_i$ ($\Tilde{\rho_1} = 0.85$, 95\% CI: 0.50, 0.99) and the correlation between $\theta_i$ and $\sigma_i$ ($\Tilde{\rho_2} = 0.24$, 95\% CI: -0.44, 0.76) and results in standardized weights close to $1/82$ for all subjects. Consequently, the BBM estimate is similar to the unweighted one. However, there are some significant fluctuations in $\lambda_i$ among the other three estimators. The weighted estimator involves assigning very large $\lambda_i$ values to the subjects with low estimated binding potential, explaining why weighting yielded smaller estimates than the other methods. The large variation in $\lambda_i$ of weighted methods is due to the large variation in $s_i$ values (range: from a low of 4.44 up through 99.52). To a lesser extent, this pattern is also seen with the trimmed estimator, though the trimming greatly reduces the range of $\lambda_i$ values. With UBM, the variation of weights is decreased even more since $\tau$ is incorporated in the weight and $\tau$ was estimated to be $25.73$ (95\% CI: 20.11, 32.84). The variation in $\lambda_i$ is further reduced in the BBM due to the large estimated value of $\rho_1$. 
\subsection{Application to meta-analysis}\label{s:meta}

    Next, we considered application of these procedures in the context of a  meta-analysis. Here, our data set contains 22 independent trials investigating the effect of selective decontamination of the digestive tract on the risk of respiratory tract infection. In all trials, patients in intensive care units were randomized to receive treatment by a combination of non-absorbable antibiotics or to receive no treatment \citep{smith1995bayesian, turner2000multilevel}. All trials reported estimates and variances for log-odds ratios (log-ORs) of respiratory tract infection between the treatment and placebo groups.

\begin{figure}
\begin{center}
\includegraphics[width=6in]{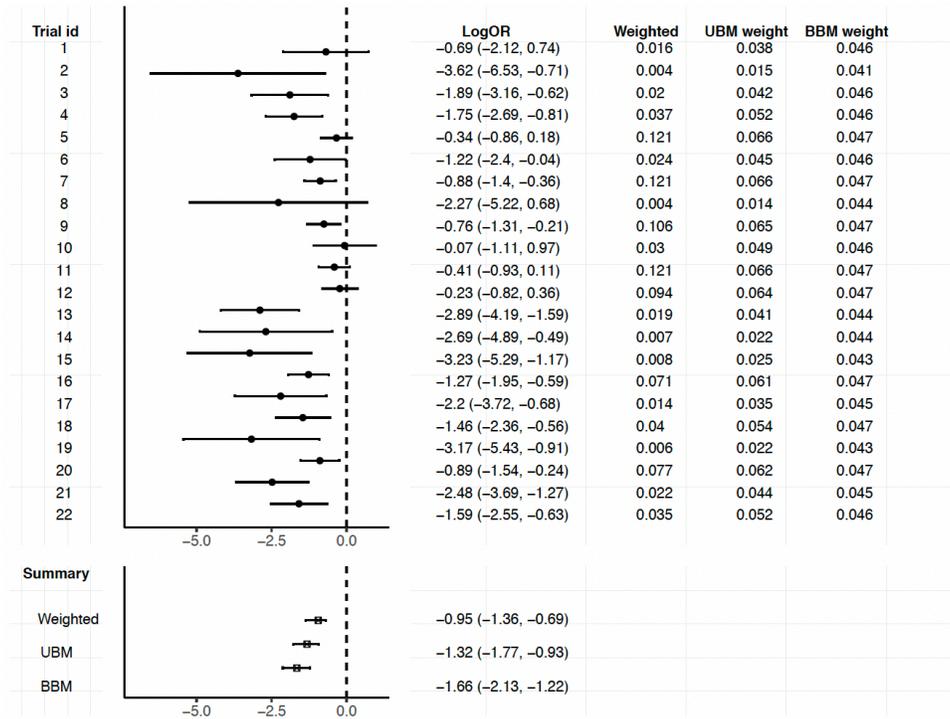}
\end{center}
\caption{Application to meta-analysis: the overall estimates and 95\% CIs for log-odds ratio of respiratory tract infection between treatment and placebo using the inverse-variance weighted, UBM and BBM methods, and comparison of impact of each trial to the overall estimate using standardized weights for each method. 
}
\label{f-11}
\end{figure}
    
    The data indicates a strong negative correlation between the summary measures and their corresponding variances (Supplementary Figure S3). We applied the BBM to the log-ORs and compared the results to the UBM and the inverse-variance weighted estimator. The BBM shows that the risk of respiratory tract infection for the treatment group is largely reduced compared to the placebo group (OR: $\exp(-1.66)=0.19$, 95\% CI: $0.12, 0.30$). The weighted estimator (OR: $0.39$, 95\% CI: $0.26, 0.50$) and UBM (OR: $0.27$, 95\% CI: $0.17, 0.39$) estimated a smaller effect than the BBM.
    Figure~\ref{f-11} shows the standardized weights of each trial data with each method for their contribution to the overall mean estimate along with the point estimates and 95\% CIs for overall mean. The BBM weights are similar across all trials; while larger inverse of variance weights are associated with trials with larger log-OR estimates (closer to zero) due to the negative association between $y_i$ and $s_i$, leading to an overall log-OR point estimate that is closer to zero and thus smaller effect estimate. The values of UBM weights fall between BBM weights and inverse of variance weights, resulting in an estimate that is smaller than the BBM but larger than the weighted estimator. The Bayesian posterior predictive $p$-value is 0.49 in the BBM model,  suggesting proper fit of the BBM model to the data (Supplementary Figure S4).

\subsection{Application to traffic safety data, small area estimation}\label{s:fars}
    Finally, we illustrated these various estimation approaches by applying them to data from a small area estimation study.  
    The Fatality Analysis Reporting System (FARS) was conducted by the National Highway Traffic Safety Administration in the United States to provide an overall measure of highway safety \citep{national2016fatality}. FARS contains data on a census of fatal vehicle crashes within the 50 states and the District of Columbia. In this application, we used 34,247 records across the 50 states and the District of Columbia from FARS 2017 to estimate the average numbers of vehicles involved in each crash in each state. This is a small area estimation problem for which states are the geographic areas of interest. 
    
    For state $i$, we can use the sample mean ($y_i$) and standard error ($s_i$) as estimates, but these can be unstable due to the sparse available data in some states. Alternatively, the Fay-Herriot model (UBM) and our proposed BBM can be applied to improve the estimation. In both models, we considered state-level covariates including resident population size, number of vehicles registered, whether the state has a law legalizing marijuana use, and geographical region.

\begin{figure}
\begin{center}
\includegraphics[width=4in]{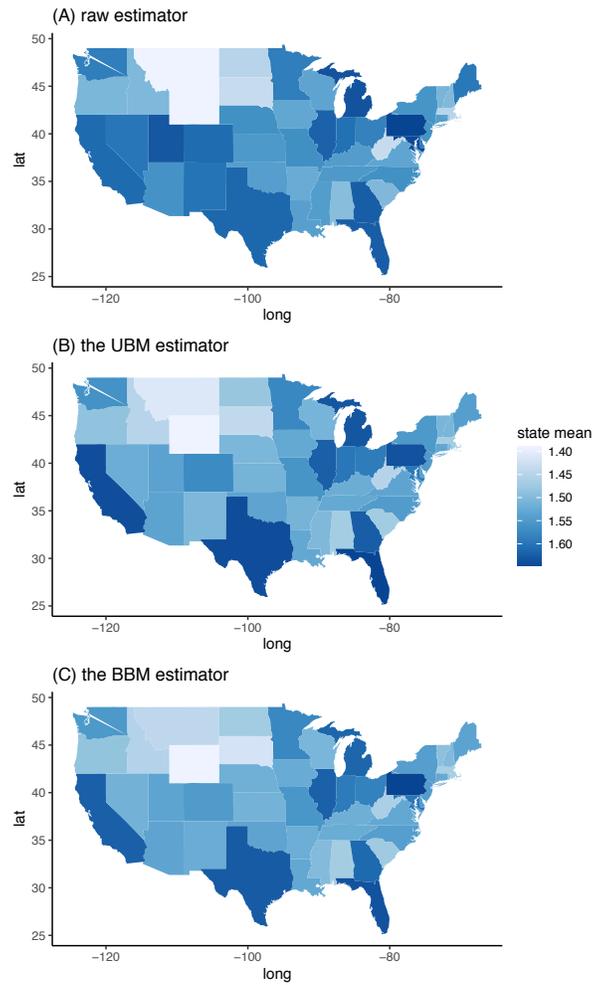}
\end{center}
\caption{Application to traffic safety data: comparison of the estimated state mean number of vehicles involved in a crash using FARS 2017 data: (A) raw estimator $y_i$ (B) the UBM estimator of $\theta_i$ (C) the BBM estimator of $\theta_i$.}
\label{f-4}
\end{figure}
      
    Figure \ref{f-4} shows maps of estimates of the mean numbers of vehicles involved in crashes for the 50 states and District of Columbia using the raw estimator $y_i$ in (A), the UBM estimator of $\theta_i$ in (B), and the BBM estimator of $\theta_i$ in (C). Supplementary Table S4 presents the point estimates of the three methods and their corresponding 95\% CIs. The raw estimates are very different to the UBM and BBM estimates in some states; while the UBM and BBM estimates are similar in most states. When they differ from the raw estimates, the BBM estimate tends to fall between the raw estimate and the UBM estimate. Compared to the raw estimate, UBM and BBM also yield shorter 95\% CIs. For example, in South Dakota, the mean is estimated to be  1.43 (95\% CI: 1.33, 1.53) by using the raw estimate;  1.49 (95\% CI: 1.41, 1.54) using UBM, and 1.47 (95\% CI: 1.41, 1.53) using BBM. Overall, the average numbers of vehicles involved in each crash in each state range fall in the range of 1.3 to 1.7, with the states of California, Florida, Georgia, Michigan, Texas and Pennsylvania reporting the highest numbers. 
        
    To better understand the differences between UBM and BBM in this application, we further examined whether $y_i$ and $\log s_i$ are correlated and whether it is appropriate to replace $\sigma_i$ with $s_i$. Supplementary Figure S5 shows the scatter plot of $\log s_i$ and $y_i$ overlaid with a loess curve, which suggests some negative correlation between the raw measure and its (log-transformed) standard error but the association is not as strong as was observed in the imaging and meta-analysis applications. After regressing on the covariates, $\rho_1$ is estimated to be 0.09 (95\% CI: -0.70, 0.80) and $\rho_2$ is estimated to be 0.03 (95\% CI: -0.65, 0.62). Both estimates are close to zero. Therefore the correlation between the measures and their uncertainty estimates is relatively weak after adjusting for the covariates. Supplementary Figure S6 shows the posterior distribution of $\sigma_i$ and compares it to $s_i$ in each state. In most states, $s_i$ falls within the 95\% CI of $\sigma_i$. 
    As expected, when $\rho_1$ is small and each $s_i$ is reasonably close to its corresponding $\sigma_i$, the UBM and BBM estimates are similar. Finally, we checked the model fit of BBM by calculating the Bayesian posterior predictive $p$-value. A $p$-value of 0.42 suggests a reasonable fit of the model (Supplementary Figure S7).

\section{Discussion} 
\label{s:discuss}

    In this paper, we propose a bivariate hierarchical Bayesian model (BBM) for combining estimates from multiple sources. This method not only models measures and measures of their uncertainty jointly, but also takes the correlation between these two quantities into consideration. 
    The simulation studies show that the BBM can provide estimates on overall mean, regression coefficients, and refined source-specific means that are less biased and more efficient with the coverage rate of 95\% CI closer to the nominal level, compared to univariate hierarchical Bayesian model (UBM) and other alternative approaches, especially if the correlation between measure and its uncertainty is not negligible. The advantage becomes more pronounced as the values of first level ($\rho_1$, observation-level) correlation and second level ($\rho_2$, population-level) correlation increase. It is interesting to note that the UBM performs poorly when $\rho_1 \neq 0$ but its performance is less sensitive to the value of $\rho_2$. Moreover, as the heterogeneity in the summary measures between data sources increases, the improvement of BBM over the alternative methods becomes more noticeable. When the variation in the summary measures between sources is small, BBM and the raw estimate perform similarly, but both still outperform the other methods. However, the raw estimator has the largest variation in estimation when the summary measures vary greatly across data sources.
    
    Our applications showed that BBM can be applied to very different data examples, with summary measures, such as mean, log odds ratio, and log rate ratio etc., and with applications in meta-analysis, small area estimation, and any other settings that combine estimates from multiple sources. We assume a bivariate normal distribution for the summary measure and its log-transformed variance estimate given the data source specific true parameter values. Transformation can be applied to the summary measure if normality assumption is not reasonable. Residuals can also be checked for the bivariate normal assumption using Q-Q plot and contour plot \citep{korkmaz2014mvn}. 
    
    Our bivariate hierarchical model for combining summary measures and their uncertainties from multiple sources is different from the bivariate meta-analysis model for sensitivity and specificity in diagnostic studies \citep{reitsma2005bivariate,chu2006bivariate}. In the bivariate meta-analysis model, two summary measures (logit sensitivity and logit specificity) are modeled jointly as a bivariate normal distribution while assuming the corresponding variance measures as fixed quantities. In contrast, our bivariate hierarchical model only considers one summary measure but assumes that the summary measure and its corresponding log-transformed standard error follow a bivariate normal distribution. For modeling two correlated summary measures, the proposed bivariate model can be extended to a multivariate model by assuming that the two summary measures and their log-transformed variance estimates follow a multivariate normal distribution with a $4\times4$ variance-covariance matrix that allows different correlations between the two summary measures and each summary measure and their corresponding variance estimate.
    
    The BBM is more computationally intensive than the UBM. The computation of the BBM is also more complex than the bivariate meta-analysis model, because the true variance parameter $\sigma_i$ for data source $i$ appears in both the mean for $\text{log}s_i$ and the residual variance for $y_i$ given $\theta_i$. To further improve computational efficiency of BBM, rather than assigning a prior distribution for $\sigma_{s_i}$, the residual variance of $\text{log}s_i$, we could take an empirical Bayes approach by setting $\sigma_{s_i}$ to be an informative fixed value, such as the empirical estimate of standard deviation (SD) of $\log s_i$. Our numerical studies show that this can greatly reduce computation time without introducing notable bias.     Therefore, in standard practice, we would recommend the BBM method with $\sigma_{s_i}$ replaced with the estimated SD of $\log s_i$. 
        
    In the applications, the estimated credible intervals obtained from BBM for the correlations are relatively wide even though the descriptive statistics shows a strong correlation. Our simulation study also shows that the credible interval gets wider as the absolute value of the correlations gets smaller. Despite the wide credible intervals for the correlation estimates, the BBM performs much better in estimating population and source-specific means than the alternative methods. Therefore, when correlations are considered as nuisance parameters in a study, BBM can provide a satisfactory result for the parameters of interest.



\clearpage


\bibliographystyle{apalike} 
\bibliography{bib.bib}      

\end{document}